\documentclass[structabstract]{aa}
\usepackage{graphicx}
\usepackage{natbib}
\usepackage{bm}
\usepackage{txfonts}
\begin{document}
   \title{MRI-driven turbulent transport: the role of dissipation, channel modes and their parasites.}

   \author{P-Y. Longaretti
          \inst{1}
          \and
          G. Lesur
          \inst{2}
          }
   \institute{Laboratoire d'Astrophysique, UJF CNRS, BP 53, 38041 Grenoble Cedex 9, France
           \and Department of Applied Mathematics and Theoretical Physics, University of Cambridge, Centre for Mathematical Sciences,
Wilberforce Road, Cambridge CB3 0WA, UK\\
              \email{Pierre-Yves.Longaretti@obs.ujf-grenoble.fr\\
}
                     }

   \date{Received XXX, year; accepted XXX}

% \abstract{}{}{}{}{}
% 5 {} token are mandatory

  \abstract
   {In the recent years, MRI-driven turbulent transport
has been found to depend in a significant way on fluid viscosity
$\nu$ and resistivity $\eta$ through the magnetic Prandtl number
$Pm=\nu/\eta$. In particular, the transport decreases with
decreasing $Pm$; if persistent at very large Reynolds numbers,
this trend may lead to question the role of MRI-turbulence in YSO
disks, whose Prandtl number is usually very small.}
  {In this context, the principle objective of the present investigation is to characterize
  in a refined way the role of dissipation. Another objective is to characterize the effect
  of linear (channel modes) and quasi-linear (parasitic modes) physics in the behavior of the
  transport.}
   {These objectives are addressed with the help of a number of incompressible numerical simulations. The
   horizontal extent of the box size has been increased in order to capture all relevant
   (fastest growing) linear and secondary parasitic unstable modes.}
   {The major results are the following:\\
   i- The increased accuracy in the computation of
   transport averages shows that the dependence of transport
   on physical dissipation exhibits two different regimes: for $Pm \lesssim 1$, the
   transport has a power-law dependence on the magnetic Reynolds number rather
   than on the Prandtl number; for $Pm > 1$, the data are consistent with a primary dependence on $Pm$ for large enough ($\sim 10^3$) Reynolds numbers.\\
   ii- The transport-dissipation correlation is not clearly or simply related to variations of the linear modes growth rates.\\
   iii- The existence of the transport-dissipation correlation
   depends neither on the number of linear modes captured in the
   simulations, nor on the effect of the parasitic modes on the
   saturation of the linear modes growth.\\
   iv- The transport is usually not dominated by axisymmetric
   (channel) modes.
   }
   {}

%test
   \maketitle
%
%________________________________________________________________

\section{Introduction}

Disks evolve on time-scales that are orders of magnitudes smaller
than expected from microphysical transport processes, and various
suggestions have been made over the years to explain this
discrepancy. Turbulent transport, in particular, has figured among
the leading candidates since the inception of the $\alpha$-disk
paradigm, and a number of hydrodynamic and MHD turbulent transport
mechanisms have been proposed in the literature.

On the hydrodynamic side, subcritical turbulence (\citealt{RZ99}
and references therein), if present, is apparently too inefficient
\citep{LL05,JBSG06}. Convection was up to now found too
inefficient and to transport angular momentum in the wrong
direction \citep{C96,SB96}, but a recent reinvestigation of the
problem indicates that this might be an artifact of these
simulations being performed too close to the stability threshold
\citep{LO10}. Two-dimensional weak turbulence
driven by small-scale, incoherent gravitational instabilities
(density waves) is an option \citep{G96}. Alternatively, the
baroclinic instability \citep{KB03} may generate vorticity, and
transport through the coupling with density waves, but its
conditions of existence are still controversial
\citep{JG06,PSJ07}, although \cite{LP10} have probably identified the
root of this debate by pointing out the nonlinear nature of the instability; 
also the resulting vortices would be subject to 3D instabilities \citep{LP09}.

\cite{BH91a} have proposed that the magnetorotational instability
(MRI) is a potentially efficient source of turbulent transport in
the nonlinear regime, an expectation soon borne out in numerical
simulations. This instability provides by now the most extensively
studied transport mechanism, through local unstratified
\citep{HGB95}, stratified \citep{SHGB96}, and global \citep{H00}
3D disk simulations. These initial simulations as well as the
numerous ones following them have shown that MRI turbulence is an
efficient way to transport angular momentum, in the presence or
absence of a mean vertical or toroidal field, with an overall
transport efficiency depending on the field configuration and
strength. However, the significant role played by microphysical
dissipation in the resolutions accessible to date had largely been
underestimated \citep{LL07,FPLH07}.

By now, both the field strength and dissipation dependence of the
simulated turbulent transport have been studied to some extent
(and only in unstratified local shearing box settings for the
latter one). The dependence of the Shakura-Sunyaev $\alpha$
parameter has been characterized very early on by \cite{HGB95} who
showed that momentum transport $\propto \beta^{-1/2}$ both for a
net vertical or toroidal field\footnote{The definition of $\beta$
(ratio of gas to magnetic field pressure) is based on the mean
field, a conserved quantity in the settings used in these
simulations.} (albeit with very different efficiencies in the two
configurations), a scaling further confirmed in later simulations,
as summarized in \cite{PCP07}.

Until recently, the effect of physical viscosity ($\nu$) and
resistivity ($\eta$) on the transport had been neglected, under
the implicit assumption that these should not matter too much once
inertial turbulent scales are resolved in the simulations.
However, \cite{LL07} have shown that, in the presence of a mean
vertical field, the MRI-driven turbulent transport did exhibit a
substantial dependence on the magnetic Prandtl number
$Pm=\nu/\eta$, with no clear trends with respect to either
viscosity of resistivity alone\footnote{The linear stability
dependence on viscosity and resistivity is totally different.}.
Recently, \cite{SH09} found similar results in shearing boxes with
a mean toroidal field instead of a mean vertical one.

When the mean magnetic flux vanishes, the transport behavior is
more complex. The initial investigation by \cite{HGB96} concluded
that the transport was converging to a finite value, but
\cite{GS05} found that the transport efficiency was dependent on
the simulation resolution. More recently, the role of the magnetic
Prandtl number $Pm$ has been identified in this setting
\citep{FPLH07}: turbulence exists only for magnetic Prandtl
numbers larger than about 2, which requires the explicit inclusion
of viscous and resistive terms in the fluid equations for
numerical simulations to correctly capture the physics of the
problem. The disappearance of turbulence at low $Pm$, as well as
the need of large enough amplitudes in the initial conditions at
$Pm > 2$, indicate that the zero net flux magnetized shearing box
is a subcritical system rather than a linearly unstable one
\citep{LO08b,LO08a}.

Thus it appears that in all configurations explored to date, the
magnetic Prandtl number plays a significant role on the existence
and/or efficiency of the turbulent transport, at least at the
resolutions accessible on present day computers (or equivalently,
the accessible Reynolds and magnetic Reynolds numbers). This
raises a number of issues.

For one, the exact role played by channel modes and parasitic
modes is unclear. Although they exist only when a mean vertical
field is present, they are simpler to analyze and their behavior
may provide insight into the generic mechanism responsible for
saturation of the linear instability. Channel modes are the
axisymmetric unstable modes of the MRI \citep{BH91a,PC08}, and are
often observed both in 2D and 3D simulations with a mean vertical
field; their name derives from their vertically layered
characteristic channel-like radial flow. They were quickly
recognized to be also nonlinear solution of the problem by
\cite{GX94}; the same authors found them to be unstable with
respect to a secondary instability (parasitic modes). A few recent
papers have focused on the possibility that the saturation of the
channel mode by this parasitic instability might be the mechanism
explaining the magnitude of the turbulent transport in MRI
simulations, with diverging conclusions \citep{PG09,LLB09}.

In relation to this, the role of the aspect ratio of the
simulations has probably been underestimated in the past. Boxes
with an aspect ratio $R:Z=1:1$ do not allow for the fastest
parasitic modes to grow, and \cite{BMCRF08} pointed out that
narrow boxes tend to overemphasize the role of the channel modes
with respect to more horizontally extended boxes. This calls for a
reassessment of the Prandtl number dependence of MRI-driven
transport in horizontally extended simulation boxes with a mean
vertical field.

More generally, it is still unclear whether this dependence of the
transport on physical dissipation is a consequence of the limited
Reynolds numbers that can be achieved on present day computers. In
particular, none of the published simulations has been able to
capture the existence of a significant inertial range in the
kinetic or magnetic energy spectrum, which makes it difficult to
address this issue. The question here revolves mostly around the
direction and locality of transfers and fluxes in Fourier space,
and will be addressed elsewhere.

For the time being, we focus the potential role of the channel and
parasitic modes in the efficiency of turbulent transport. This is
explored by numerical simulations in the shearing sheet limit,
with a net vertical magnetic flux, and with horizontally extended
simulation boxes. The paper is organized in the following way. Our
numerical method, setup, and run parameters are described in
section 2. Relevant aspects of the theory of channel 
modes are summarized in section 3. Section 4 is the core of this
paper, and discusses our numerical results; the issues bearing on
the resolved linear and secondary modes are also discussed there.
The implications of these results are presented in the final
section along with some possible future lines of work.

\section{Numerical model}

\subsection{Shearing box model and equations:}

Following the initial investigation of 3D MRI turbulent properties
by \cite{HGB95}, we base our simulations on the shearing sheet
local approximation and the related shearing box model. Most if
not all local studies of disk turbulence have been performed in
this framework. Local simulations are unescapable to examine in
any detail the structure and transport properties of MRI
turbulence; indeed, even within a local model, present day
computers are still too limited to reach the resolutions required
to understand the magnetic Prandtl issue summarized in the
introduction, and it is certainly hopeless to tackle this problem
directly in global simulations.

To some extent, a shearing box biases the role of the channel mode
in turbulent transport, e.g. through the correlations introduced
by the periodic boundary conditions. Note however, that, in purely
hydrodynamic turbulence, the shearing box seems to capture some of
the correct physical properties of actual experimental systems,
such as the transition Reynolds number to turbulence as a function
of rotation (see, \citealt{LL05}). In any case, it is very
difficult to formulate a well-posed local problem that does not
rely on the shearing box framework. The reader is referred to
\cite{HGB95}, \cite{B03} and \cite{RU08} for more detailed
discussions of the properties and limitations of this model.

MHD turbulence in discs is essentially subsonic, and we will work
in the incompressible approximation, which allows us to eliminate
local and transient density fluctuations as well as sound waves
and density waves from the problem. Density waves are excited in
shearing box turbulence \citep{HP08b}, but they appear to have
little impact on the turbulent transport \citep{HP08a}. This has been confirmed by direct
comparison between incompressible and compressible simulations
\citep{FPLH07}. As a consequence, we feel reasonably justified to
assume incompressibility. Explicit molecular viscosity and
resistivity are included.

The shearing box equations are well-known; we reproduce them here
to introduce our notations. We chose a Cartesian box centered at
$r=R_0$, rotating with the disc at angular velocity
$\Omega=\Omega(R_0)$ and having dimensions $(L_x,L_y,L_z)$ with
$L_i \ll R_0$. By convention here, $R_0\phi = x$ and $r-R_0 = -y$,
leading to the following form of the  shearing sheet equations:

\begin{eqnarray}
\nonumber \partial_t \bm{U}+\bm{\nabla\cdot} (\bm{U\otimes U})&=&-\bm{\nabla} \Pi+\bm{\nabla\cdot }(\bm{ B \otimes B}) \\
& &-2\bm{\Omega \times U}+2\Omega S y \bm{e_y}+\nu \bm{\Delta U},\\
\partial_t \bm{B}&=&\bm{\nabla \times} (\bm{U \times B}) +\eta \bm{\Delta B},\\
\label{divv} \bm{\nabla \cdot U}&=&0,\\
\bm{\nabla \cdot B}&=&0.
\end{eqnarray}

\noindent where the magnetic field is measured in units of
Alfv\'en speed. The mean shear $S=-r\partial_r \Omega$ is set to
a Keplerian flow value $S=(3/2)\Omega$. The generalized pressure term
$\Pi$ includes both the gas pressure term $P/\rho_0$ and the
magnetic one $\bm{B}^2/2\rho_0$. This generalized pressure $\Pi$
is fixed by the incompressibility condition Eq.~(\ref{divv}), and
computed by solving a Poisson equation. The magnetic field is
expressed in Alfv\' en-speed units, for simplicity.

The steady-state solution to these equations is the local
Keplerian profile $\bm{U}_0=Sy\bm{e_x}$. Our code computes the
(turbulent) deviations from this Keplerian profile. Defining
$\bm{v}=\bm{U}-\bm{U}_0$, and $\bm{b}=\bm{B}-\bm{B}_0$, one
obtains the following equations for $\bm{v}$ and $\bm{b}$:

\begin{eqnarray}
\nonumber \partial _t \bm{v}&=&
-\bm{v}\cdot\bm{\nabla}\bm{v}-\bm{\nabla} \Pi +
\bm{B}\cdot\bm{\nabla}\bm{b}-Sy\partial_x \bm{v}\\
\label{motion}& & +(2\Omega-S) v_y\bm{e_x}-2\Omega v_x \bm{e_y}+\nu\bm{\Delta v},\\
 \nonumber \partial _t \bm{b}&=&-\bm{v}\cdot\bm{\nabla}\bm{b}+\bm{B}\cdot\bm{\nabla}\bm{v}\\
 \label{induction}& & -Sy\partial_x \bm{b}+S b_y\bm{e_x}+\eta \bm{\Delta b},\\
\label{Vstruct} \bm{\nabla \cdot v}&=&0,\\
\label{Bstruct}\bm{\nabla \cdot b}&=&0.
\end{eqnarray}

The boundary conditions associated with this system are periodic
in the $x$ and $z$ direction and shearing-periodic in the $y$
direction \citep{HGB95} (for $\bm{v}$ and $\bm{b}$).

Following \cite{HGB95}, one can integrate the induction equation
(\ref{induction}) over the volume of the box, leading to:

\begin{equation}
\label{consB} \frac{\partial \langle \bm{B}\rangle }{\partial
t}=S\langle B_y\ \rangle \bm{e}_x,
\end{equation}

\noindent where $\langle \rangle$ denotes a volume average.
Therefore, the mean magnetic field is conserved, provided that no
mean radial field is present. In this work, a mean vertical field
$B_0$ is imposed, and conserved by virtue of Eq.~(\ref{consB}).

The numerical resolution makes use of a spectral Galerkin
representation of equations (\ref{motion})--(\ref{Bstruct}) in the
sheared frame \citep[see][]{LL05}. In this frame, the
shearing-sheet boundary conditions are transformed into perfectly
periodic boundary conditions, and Fourier transforms can be used
in all three directions. Moreover, this decomposition allows us to
conserve magnetic flux to machine precision (the total magnetic
flux created during one simulation is typically $10^{-11}$).
Equations (\ref{Vstruct}) and (\ref{Bstruct}) are enforced to
machine precision using a spectral projection \citep{P02}. The
nonlinear terms are computed with a pseudospectral method, and
aliasing is prevented using the 3/2 rule.

The sheared frame representation of the spectral domain would
eventually produce a mismatch between the computed Fourier
components and the physically relevant ones. To circumvent this
problem, we use a remap method similar to the one described by
\cite{UR04}. This routine redefines the sheared frame every
$T_{\mathrm{remap}}=L_x/(L_yS)$, and none of the
results presented here seems to be related to this time scale.
Spectral methods are very little dissipative by nature; numerical
dissipation is kept to very small values, as can be seen by
computing the total energy budget at each time step (see
\citealt{LL05} for a discussion of this procedure).

\subsection{Dimensionless numbers:}\label{dimnum}

All our simulations are performed in horizontally extended boxes,
with aspect ratio $L_x:L_y:L_z = 4:4:1$, with a non vanishing mean
vertical field of varying strength.

Our shearing box set up is characterized by a number of
dimensionless numbers. Our simulations explore the dependence of
the turbulent transport with respect to three of them

\begin{itemize}
\item The magnitude of the imposed mean vertical field measured by
\begin{equation}\label{beta}
\beta=\frac{S^2 L_z^2}{V_A^2},
\end{equation}
where is the Alfv\'en speed due to this mean field. This
definition mimics the usual plasma $\beta$ in vertically
stratified disk obeying the vertical hydrostatic equilibrium
constraint $c_s \sim \Omega
  L_z$.
\item The Reynolds number,
\begin{equation}\label{Re}
Re=\frac{SL_z^2}{\nu},
\end{equation}
comparing the nonlinear advection term to the viscous dissipation.
\item The magnetic Reynolds number,
\begin{equation}\label{Rm}
Rm=\frac{SL_z^2}{\eta},
\end{equation}
comparing magnetic field advection to the Ohmic resistivity.
\item Alternatively to one of the two Reynolds numbers, the magnetic Prandtl
number
\begin{equation}\label{Pm}
Pm = \frac{Rm}{Re} = \frac{\nu}{\eta}.
\end{equation}
\end{itemize}

Our study focuses on the dependence of the turbulent transport on
$\beta$, $Re$ and $Rm$, but other combinations of the Reynolds
numbers will be used when necessary. Note however that
unstratified shearing boxes are also characterized by two other
numbers that are constant for our simulations:

\begin{itemize}
  \item The Mach number
\begin{equation}\label{Ma}
Ma = \frac{S L_z}{c_s}\ (= 0),
\end{equation}
(vanishing for our incompressible
  simulations)
  \item A Rossby-like number $q$ defined by
\begin{equation}\label{Romega}
q = -\frac{S}{\Omega},
\end{equation}
which is fixed at the keplerian value (3/2). The sign of $q$ indicates the flow cyclonicity.
\end{itemize}

In the following we use $S^{-1}$ as the unit of time and $SL_z$ as
the unit of velocity. One orbit corresponds to
$T_\mathrm{orb}=3\pi S^{-1}\simeq 10 S^{-1}$; our simulation
lengths are to be divided by about 10 to compare them with
simulations scaled in orbit times, as is commonly done by
other workers in the field. For simplicity, in what follows, we
keep the same notation for dimensionless and dimensional
quantities.

Our dimensionless transport coefficient $\alpha$ is defined as

\begin{equation}\label{alpha-def}
  \alpha=\frac{\langle b_x b_y - v_x v_y \rangle}{S^2 L_z^2},
\end{equation}

\noindent where the average is taken over the volume of the
simulation box. Within a factor of order unity, this is analogous
to the more common prescription for the turbulent viscosity
$\nu_t=\alpha c_s H$ if one identifies $L_z$ with the disc thickness
$H$. 

\section{Channel mode physics: a summary of relevant
information:}\label{chan-paras}

As recalled in the introduction, it has been noted in a number of
earlier MRI simulations in a shearing box with a mean vertical
field that channel modes constitute somewhat recurrent patterns of
the turbulent flow, and appear to be more prominent at the maxima
in the fluctuations of the turbulent transport. Channel modes do
transport angular momentum with roughly the right order of
magnitude if their amplitude is comparable to the rms fluctuation
in the field close to a maximum of the transport.

\begin{figure}[ht]
   \centering
   \includegraphics[scale=0.35]{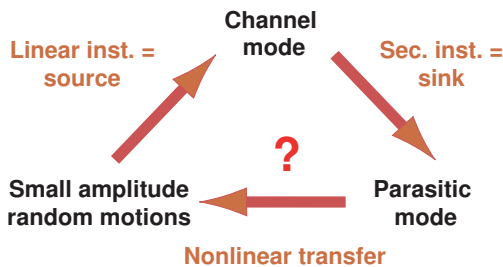}
   \caption{Cartoon of the process that may be at the origin of the
   saturation of the transport in shearing box simulations of MRI with
    a net vertical field (see text).}\label{satproc}
\end{figure}

This has suggested a picture of turbulent transport in shearing
boxes with a mean vertical field that is sketched on
Fig.~\ref{satproc}.

According to this picture, the channel modes linearly grow from
random noise; at some amplitude, their growth is halted by a
secondary instability (parasitic mode) which destroys the channel
mode. Presumably, the parasitic modes themselves decay into small
scale turbulence, which may then produce the seed for the random
fluctuations out of which the channel mode grows in the first
place. In this scenario, the channel mode(s) would be responsible for most of the transport, in particular near maxima of the transport fluctuations.

One of the objectives of this paper is to examine the relevance of
this picture of MRI turbulence. Even if such a scenario is not
generic (it depends directly on the presence of a mean vertical
field), if confirmed, it might provide an interesting lead to
analyze different situations. Another related objective it to
assess the relevance of this scenario to the question of the
Prandtl number dependence of MRI-driven transport. To this effect,
we gather here the relevant pieces of information on the physics
of channel modes that is required in order to
analyze the simulations presented in the next section.

\subsection{Dispersion relation}\label{rel-disp}  

The physics of viscous and resistive channel modes has been
examined in \cite{LL07} and their properties have been
characterized in detail in \cite{PC08} (\citealt{SM99} have also
explored the role of resistivity on MRI in the absence of
viscosity). Since non axisymmetric MRI modes are
transiently growing structures in non ideal MHD \citep{BH92},
 the axisymmetric modes give a good grasp of the
linear stability properties of the shearing box.

\begin{figure}[ht]
   \centering
   \includegraphics[scale=0.6]{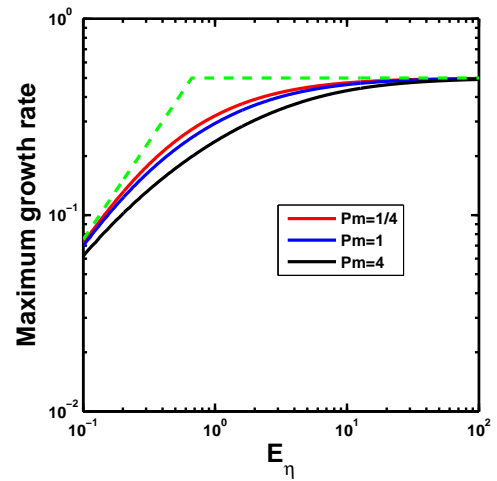}
   \caption{Maximal growth rate $\sigma_M/S$ as a function of the resistive dissipation, for
   the various values of the Prandtl number used in the simulations. Also shown
   are asymptotic analytic approximations (see text).}\label{growthrate}
\end{figure}

We recall here the dispersion relation of theses modes and a
number of other features that will be of use in the next section.

Looking for solutions of the linearized equations of motions in
the form $\bm{v}=\bm{v_l}\exp(\sigma t - i k_y y - i k z)$ and
$\bm{b}=\bm{b_l}\exp(\sigma t - i k z)$ leads to the following
fourth order dispersion relation:

\begin{eqnarray}
\nonumber \sigma^4&+2k^2\sigma^3(\eta+\nu)+\sigma^2\Big(a+k^4(\eta^2+\nu^2+4\eta\nu)+b\Big)\\
\nonumber &+\sigma\Big(2k^6(\eta\nu^2+\nu\eta^2)+ak^2(\nu+\eta)+2b\eta k^2\Big)\\
\label{disp_rel} &+\nu^2\eta^2k^8+a\nu\eta k^4+b\eta^2k^4-c=0,
\end{eqnarray}

\noindent with

\begin{eqnarray}
a&=&2k_z^2V_A^2,\\
b&=&\kappa^2 \gamma^2,\\
c&=&k_z^2 V_A^2(2\Omega S \gamma^2 - k_z^2 V_A^2),
\end{eqnarray}

\noindent and where $\kappa=[2\Omega(2\Omega - S)]^{1/2}$ is the
epicyclic frequency, $V_A = B_0$ is the Alfv\'en speed based on
the imposed mean vertical field\footnote{We measure the magnetic
field in units of Alfv\'en speed.}, and
$\gamma^2=k_z^2/(k_y^2+k_z^2)$. Channel modes have $k_y=0$ so that
$\gamma=1$.

\begin{figure*}[ht]
   \centering
   \includegraphics[width=0.7\linewidth,angle=-90]{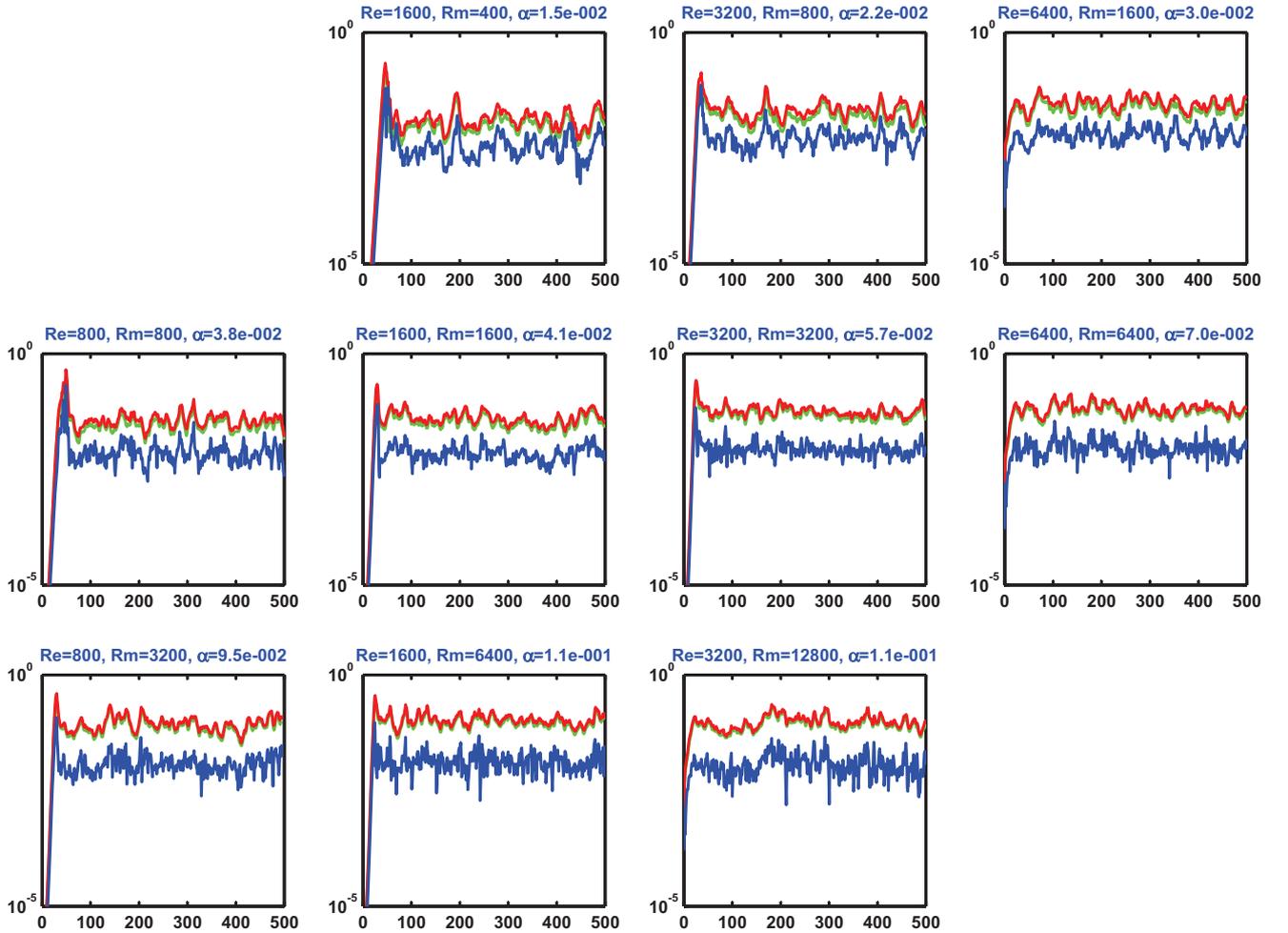}
   \caption{The dimensionless turbulent transport $\alpha$ as a function of time (in units of $S^{-1}$) for a vertical
   magnetic field strength $\beta=1000$ and various levels of viscous
   and resistive dissipation $Re$ and $Rm$. Insets on a given line correspond to constant
   magnetic Prandtl number, $Pm= 1/4$, $1$ and $4$ from top to bottom. The
   Reynolds number $Re$ is constant on rows ($800$, $1600$, $3200$, and $6400$
   respectively. The transport averages are computed over the last $400$ shear times
   of the simulations. The blue line represents the Reynolds stress, the green line
   the Maxwell stress, and the red line the total stress (these last two are almost indistinguishable).}\label{ttransp}
\end{figure*}

This equation is used later on to compute the channel mode growth
rates in the conditions of our numerical simulations. It is most
conveniently solved in dimensionless form, with $S$ as unit of
time and $V_A/S$ as unit of length. We introduce
$\sigma^*=\sigma/S$, $\kappa^*=\kappa/S$, $k^*=V_A k/S$; defining
the viscous ($E_\nu$) and resistive ($E_\eta$) Elsasser-like
numbers by

\begin{eqnarray}
E_\nu&=&\frac{V_A^2}{S\nu}=\frac{Re}{\beta}=\frac{\Lambda_\nu}{q},\label{elsasnu}\\
E_\eta&=&\frac{V_A^2}{S\eta}=\frac{Rm}{\beta}=\frac{\Lambda_\eta}{q},\label{elsaseta}
\end{eqnarray}

\noindent one has $\nu^*=1/E_\nu$ and $\eta^*=1/E_\eta$. Note that
with a power-law velocity profile in a disk ($\Omega\propto
r^{-q}$), $\Omega^*=1/q$ and $\kappa^*=[2(2-q)]^{1/2}/q$. Our
definitions of the Elsasser numbers (and other dimensionless
quantities) differ from the usual ones ($\Lambda_\nu$ and
$\Lambda_\eta$, see e.g. \citealt{PC08} and \citealt{PG09}) by a
factor $S/\Omega$. This choice has several motivations: $S$ rather
than $\Omega$ is the relevant time-scale of linear growth rates;
this definition is more consistent with our previous choice of
units; and it makes the connection between the Elsasser and
Reynolds numbers simpler.

\subsection{Stability limits}

In the dissipation-free limit, the dispersion relation can be
solved exactly. The magnetic tension stabilizes the instability
for $k^* > k_c^*\equiv (2\Omega^*)^{1/2}=(2/q)^{1/2}$. In a box of
finite vertical extent $L_z$, the vertical modes wavelengths are
multiple of $k_{min}=2\pi/L_z$. Axisymmetric MRI modes are
therefore stabilized when $k_c < k_{min}$, which
translates into $\beta \ge \beta_c\equiv 2\pi^2 S/\Omega\simeq
30$. The maximum growth rate in the dissipation-free regime,
$\sigma_M=S/2$, is achieved for a wavenumber $k_M\simeq k_c/2$.

Resistivity and viscosity modify the marginal stability limit in
the ($Re$, $Rm$, $\beta$) space. Numerical investigation indicates
that only one of the four roots is unstable for $q < 2$, so that
the marginal stability limits obtains when the
$\sigma$-independent term of the dispersion relation is equal to
zero. Also, as shown by \cite{LL07}, for $\beta \ge 100$ and
$Re\gtrsim 400$ (two conditions that are satisfied in the present
investigation), the marginal stability limit is nearly independent
of the Reynolds number. This leads to the following expression of
the marginal stability limit

\begin{equation}\label{etalim}
Rm=\frac{\kappa\beta}{S}\left(\frac{\beta_c}{\beta-\beta_c}\right)^{1/2}=
\frac{2\pi}{\sqrt{3}}\frac{\beta}{(\beta-\beta_c)^{1/2}}\sim
3\beta^{1/2},
\end{equation}

\noindent where the second equality applies for keplerian flows.
The last approximation holds only when $\beta$ is large enough,
and was given in \cite{LL07}; the first expression is new and
sensibly more precise.

The maximum growth rate achieved by the instability is modified
when viscosity and resistivity become important, i.e. when either
$E_\nu$ or $E_\eta$ $\lesssim 1$. \cite{PC08} give an asymptotic
form of the growth rate in the dissipation regime for small or of
order unity $Pm$ (see their Fig.~5 and their Eq.~91); it reads
$\sigma_M=2\sigma_0 E_\eta/3$ where $\sigma_0=S/2$ is the
dissipation-free growth rate. The range of $E_\eta$ and $Pm$
examined in this work spans the transition from a dissipative to a
non dissipative regime for the linear instability. The
corresponding variation of maximum growth rate with both $E_\eta$
and $E_\nu$ (through the Prandtl number) is shown on
Fig.~\ref{growthrate}, along with the asymptotic expressions just
recalled.

\begin{figure}
   \centering
   \includegraphics[scale=0.5]{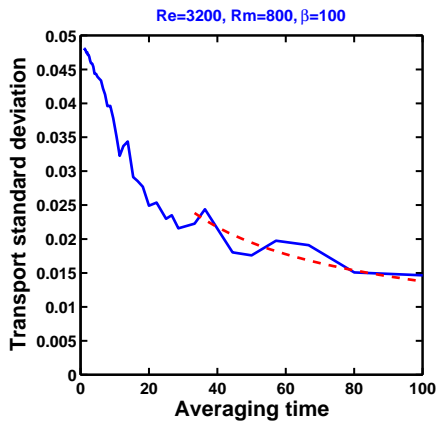}
    \caption{Transport standard deviation as a function of the time interval $\tau$ used to bin the transport
   data, for a given run (in this case, $Re=3200$, $Rm=800$, and $\beta=100$). This information is used to
   quantify the error in the transport from the fit $\propto \tau^{-1/2}$ that is expected to hold for large enough
   binning time $\tau$ (see text for details).}
              \label{alphadev}%
\end{figure}

\section{Numerical results}

In order to explore a significant range of parameters, most of our
runs are performed at a standard resolution ($N_r$, $N_\phi$,
$N_z$) = (256, 128, 64) in real space. All simulations have an
aspect ratio $R:\phi:z=4:4:1$; as argued earlier, this aspect
ratio allows us to capture the fastest growing channel and
parasitic modes. Three simulations have been performed at a
resolution twice as large (512, 256, 128) in order to reach a
higher Reynolds number ($Re=20000)$, and lower Prandtl numbers
(down to $Pm=1/16$). Note that spectral codes are intrinsically
more resolved than finite difference codes with the same number of
points; as a consequence, one needs to adopt resolutions larger by
a factor $\simeq 2$ in all directions in a finite difference code
such as ZEUS, Athena or RAMSES to obtain the same results, a point
to bear in mind when comparing our conclusions with those published in
the literature with finite difference codes \citep[see e.g.][for
an explicit comparison]{FPLH07}.

\begin{figure}
   \centering
   \includegraphics[scale=0.5]{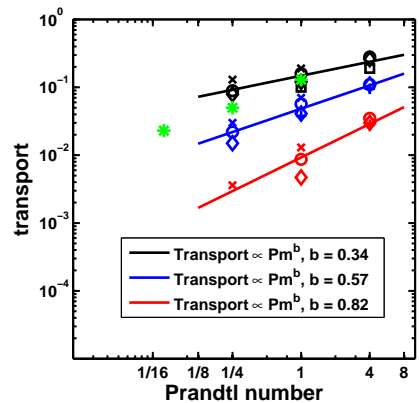}
   \caption{The dimensionless turbulent radial momentum transport
   $\alpha$ as a function of the Prandtl number $Pm$ and the field
   strength $\beta$ for various Reynolds numbers. black:
   $\beta=10^2$; blue: $\beta=10^3$; red: $\beta=10^4$; $\Box$: $Re=400$; +: $Re=800$;
   $\diamondsuit$: $Re=1600$; $\circ$: $Re=3200$;
   $\times$: $Re=6400$; the green starred data points correspond
   to the three more resolved runs at $Re=20000$ and $\beta=10^3$. 
   Power law fits are also shown for each value of $\beta$.}
\label{transp-pm}
\end{figure}

The simulations performed in this work are in the large Reynolds
number limit ($400 < Re < 20000$). The relevant marginal stability
limit depends only on $Rm$ and is precisely given by
Eq.~(\ref{etalim}). When the magnetic Reynolds number is too close
to this limit, nonaxisymmetric perturbations damp
out, and the flow becomes nearly identical to an axisymmetric one. We have checked that none of the simulations discussed below is affected by this issue.

\begin{figure*}[!ht]
   \centering
   \includegraphics[scale=0.5]{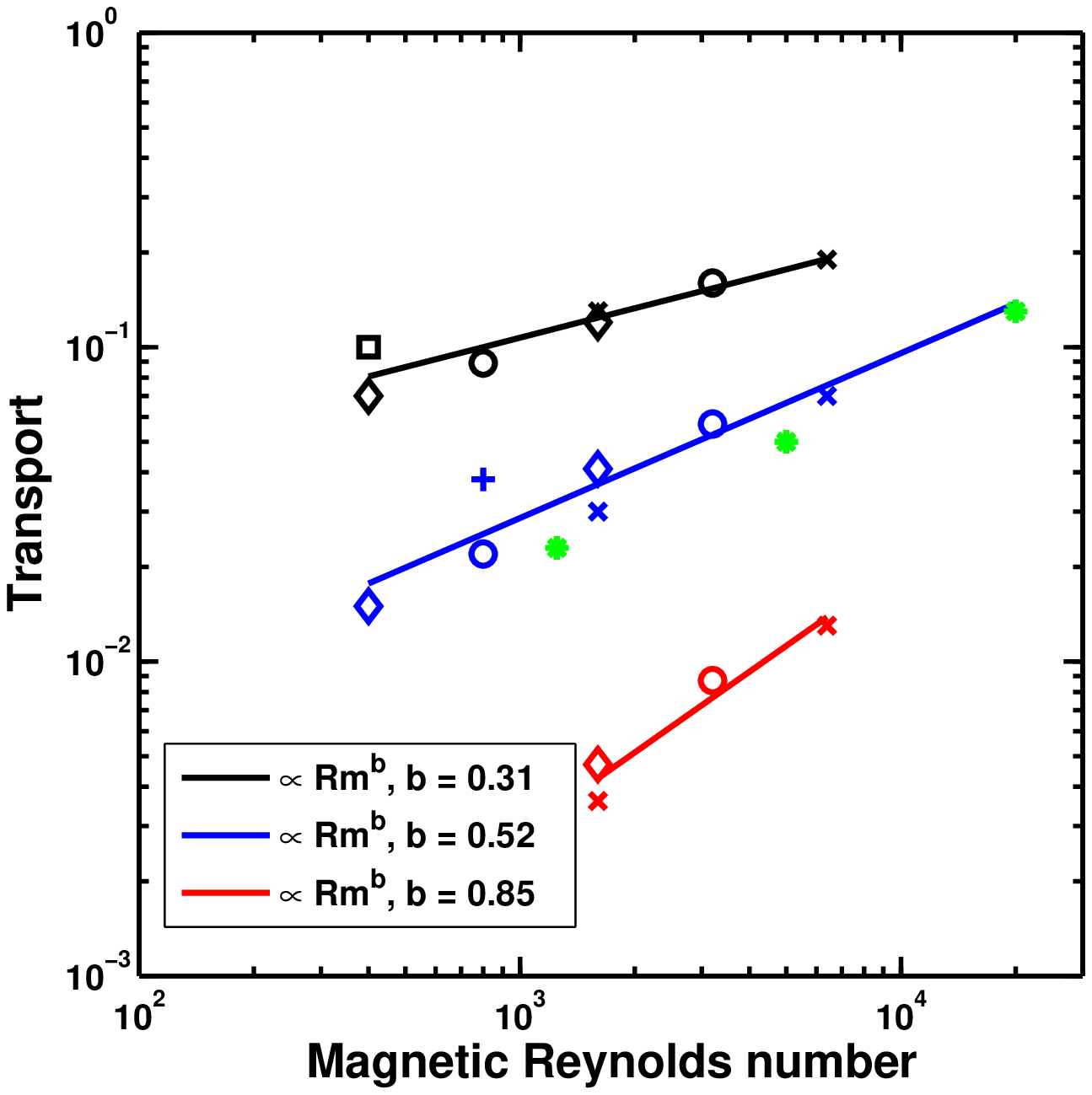}
   \includegraphics[scale=0.5]{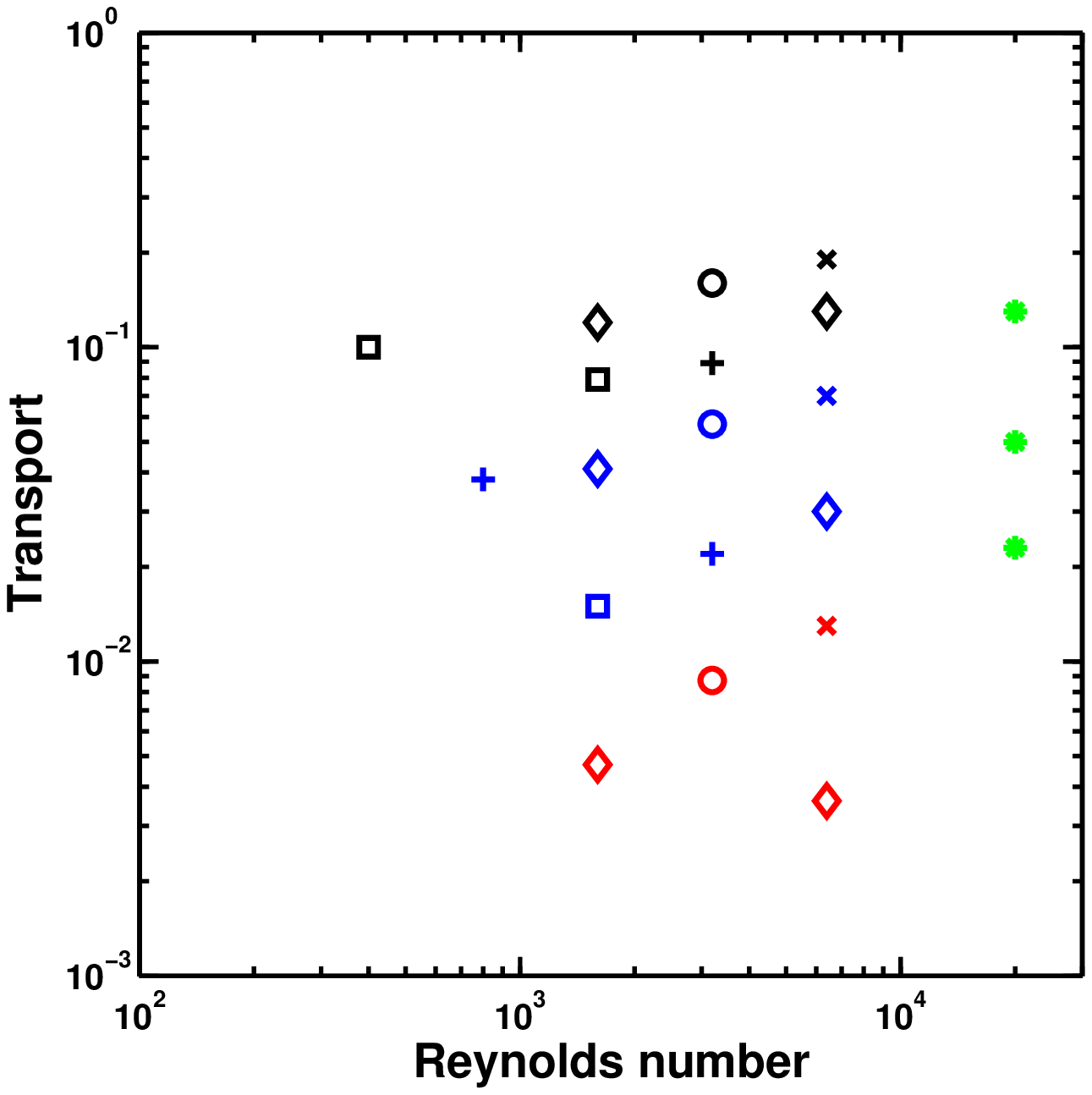}
   \caption{The dimensionless transport $\alpha$ as a function of the
   magnetic Reynolds number $Rm$ (left) and Reynolds number $Re$ (right).
   Black: $\beta=10^2$; blue: $\beta=10^3$; red: $\beta=10^4$; $\Box$: $R=400$; +: $R=800$;
   $\diamondsuit$: $R=1600$; $\circ$: $R=3200$; $\times$: $R=6400$; $R$ refers
   to $Re$ on the left panel, and to $Rm$ on the right one. The green stars correspond to the three more
   resolved runs at $Re=20000$
   (see text).}
              \label{transpReRm}%
\end{figure*}

With our adopted standard resolution, we can confidently resolve
the dissipative scales up to Reynolds numbers $\simeq 10^4$.
Consequently, we have explored Reynolds numbers in the range $400$
-- $6400$ and magnetic Reynolds numbers in the range $400$ --
$12800$. The magnetic Prandtl number of the simulations is either
$1/4$, $1$ or $4$. A smaller value ($Pm=1/16$) has also been
reached with one of our higher resolution runs with $Re=20000$.

We have performed simulations for three different vertical field
strengths: $\beta=10^2$, $10^3$, and $10^4$. This range is
constrained by two considerations. On the low end (large field),
one must stay away from the marginal stability limit due to the
magnetic tension. Indeed, when the incompressible MRI works too
close to this stability threshold, the flow is strongly biased by
the presence of channel modes (an effect that incompressible
simulations are more prone to capture, as exemplified in
\citealt{LL07}). On the large end (weak field), one is limited by
the fact that at some point, the flow is going to be dominated by
the zero flux transport process, whose dependence on the viscous
and resistive dissipation might be sensibly different. If one
simply assumes that one shifts from one regime to the other when
the efficiency of the transport processes are equal, one finds
that the transition between the mean field and zero net field
regimes should be in the range $\beta\sim 10^{5}$ -- $10^{4}$;
this limit applies to $Pm \gtrsim 2$ (as the zero net flux
transport vanishes for lower Prandtl number values), and should be
somewhat dependent on $Pm$. As discussed later on, for our
weakest field ($\beta = 10^4$) and smallest Prandtl ($Pm=4$), the
flow also tends to become bidimensional for low enough Reynolds
number. 

\subsection{Transport time histories and standard deviation}

Figure \ref{ttransp} displays the behavior of the turbulent
transport for the runs we have performed at $\beta=1000$. The
simulations are typically run for $500$ shear times, and the
transport average is based on the last $400$. In our horizontally
extended simulation boxes, the transport fluctuations are
substantially reduced with respect to narrower boxes. Typically,
fluctuations of a factor of $\sim2$ are observed, whereas in $r:z =
1:1$ boxes, fluctuations of an order of magnitude or more are
common. This reduction provides more precise averages for the
transport, even though the averaging time is somewhat smaller than
what is used for boxes with a $R:Z=1:1$ aspect ratio. This aspect ratio dependence is related to the
prominence of channel modes in the large transport bursts observed
in the narrow boxes \citep{BMCRF08}, a feature related to the fact that width of the narrow box is comparable to the
correlation length of turbulent fluctuations in the horizontal direction \citep{GGSJ09}.

The transport time history can be used to quantify the error in
the determination of $\alpha$ in the following way. From the raw
time history, one can define a series of binned time histories,
with a binning time $\tau$ ranging from $1$ to $100$ shear
times\footnote{For $\tau < S^{-1}$, the transport statistics
depends very little on the binning time, while for $\tau > 100
S^{-1}$, the statistics is too poor to be meaningful.}. From these
binned time histories, one can define a transport standard
deviation in the usual way, i.e.,
$\sigma_\alpha=[\sum(\alpha_i-\overline\alpha)^2/N]^{1/2}$ where
$N$ is the number of bins, $\alpha_i$ itself being the average
value of the transport in bin $i$. The resulting dependence
$\sigma_\alpha(\tau)$ is shown on Fig.~\ref{alphadev} for one of
our runs. Two regimes can be distinguished. For $\tau \lesssim 10$
-- $20\ S^{-1}$, the deviation decreases sharply with $\tau$; this
is seen directly on Fig.~\ref{ttransp}, where the transport
typical variation time scale is precisely in this range. For large
values of $\tau$, the transport is less and less correlated with
itself. Consequently, one expects that it should behave more or
less as a random walk, so that the standard deviation should scale
like $\tau^{-1/2}$. A fit of this type is performed on
Fig.~\ref{alphadev}, and appears to represent reasonably well the
trend, in spite of the crudeness of the approximation.

From this approximation we find that the relative error in the
computation of the transport over $400$ shear times is in the
range $3$ -- $10\%$. There is no clear tendency for the larger
values of the relative error to correspond to higher transport,
and no clear trend with either the magnetic Prandtl or magnetic
Reynolds number. On average, $\sigma_\alpha/\alpha\sim 5\%$.

\subsection{Physical dissipation and transport}

The dependence of the transport on the dissipation for various
field strengths is presented on Fig.~{\ref{transp-pm} for all our
standard resolution runs. This figure significantly extends the
results reported in \citep{LL07}. Several trends can be observed:

\begin{itemize}
  \item For a given field strength, the dependence on the magnetic
Prandtl number $Pm$ seems to follow an approximate power law. 
For comparison with our previous work, power-law fits are computed. 
The exponents are $\simeq 0.3$, $0.7$ and $0.8$ for the three values 
of $\beta$, in increasing order.
 \item The three more resolved runs confirm this trend: the
  transport is substantially larger than other runs at the same
  field strength ($\beta=10^3$), but the $Pm$ dependence is consistent
  with the other runs at the same $\beta$;
  this $Pm$ dependence does not seem to saturate even though a lower
  Prandtl number has been achieved (1/16).
  \item At a given Prandtl number, the transport is decreasing
with decreasing field strength (increasing $\beta$). This trend is
expected from the scaling $\alpha\propto \beta^{-1/2}$ that was
first pointed out in the initial work of \cite{HGB95}. However,
the variation of the index of the Prandtl number dependence with
field strength just discussed induces deviations from this
scaling.
  \item There is a weaker, but systematic and significant increase
of the transport with increasing Reynolds number at any given
field strength and Prandtl number. This effect is real:
for most of our runs, this increase is larger than the standard
deviation in the transport, as quantified in the previous
subsection. It is also larger for the smaller Prandtl number values.
This indicates that the Prandtl number does not capture all the physics of the
correlation between transport and physical dissipation; this point is 
further discussed below.
  \end{itemize}

Our previous investigation was limited to $\beta=100$. In the
present work, the Prandtl number dependence of the
transport for this field strength is consistent with our earlier
findings. However, the transport observed here is reduced by a
factor $\sim 2$; this is a direct consequence of the reduced role
played by the channel mode in our horizontally extended simulation
boxes.

\begin{figure*}[!ht]
   \centering
   \includegraphics[width=0.75\linewidth]{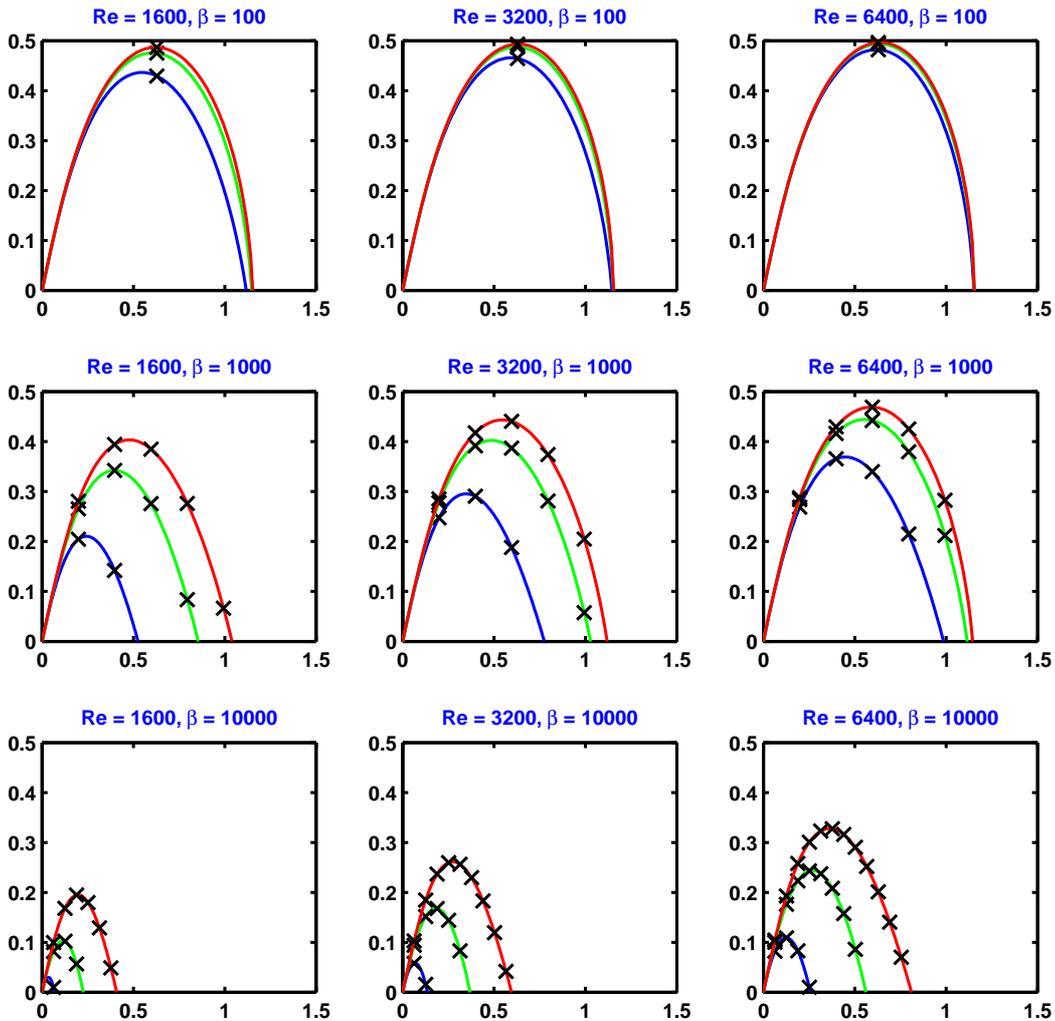}
   \caption{Composite figures showing the dimensionless growth rate $\sigma/S$ of the channel mode for most of the
   various runs performed, as a function of the dimensionless wavenumber $V_A k_z/S$. Each subplot is labelled by its
   Reynolds number and field strength ($\beta$). The various
   curves in each subplot correspond to $Pm=1/4$, $1$, $4$,
   respectively (the growth rate increases with increasing Prandtl
   number). The crosses indicate the modes that are represented in
   our simulation boxes, as a result of the boundary conditions (see text for a
   discussion).}\label{growthrate-comp}
\end{figure*}

For the lowest Prandtl number ($1/4$) and lowest field strength
($\beta=10^4$), only one point is reported in the graph. Our other
runs for this parameter have lower Reynolds numbers, and are too
close to the linear stability threshold of Eq.~(\ref{etalim}) to
sustain full 3D turbulent motions. These runs show 2D or quasi-2D
behavior, with very different transport efficiency and behavior.
The data point we have retained might still be weakly affected by
such effects.

As pointed out above, Fig.~\ref{transp-pm} indicates that the spread with
Reynolds number at a given Prandtl number increases with
decreasing Prandtl number. In fact, two different regimes can be
noted, one for $Pm \le 1$ and one for $Pm =4$.

At $Pm=4$, the transport seems to be only weakly dependent on $Re$ (or $Rm$), at least for large enough Reynolds number: the transport increases by $10\%$ to $50\%$ (depending on $\beta$)
while the Reynolds number is multiplied by a factor of 4. This trend can also be found in the work of \cite{SH09}, where the MRI turbulent transport in presence of a toroidal field is investigated with more emphasis on the $Pm > 1$ regime. Their Figs.~6 and 7 show that, for $Pm=2$ and $4$ at least (the only ones with enough data in the $Pm > 1$ regime), the transport increases steadily with the Reynolds number for $Re \lesssim 1000$ and much more weakly for $Re \gtrsim 1000$.

\begin{figure*}[ht]
   \centering
   \includegraphics[scale=0.5]{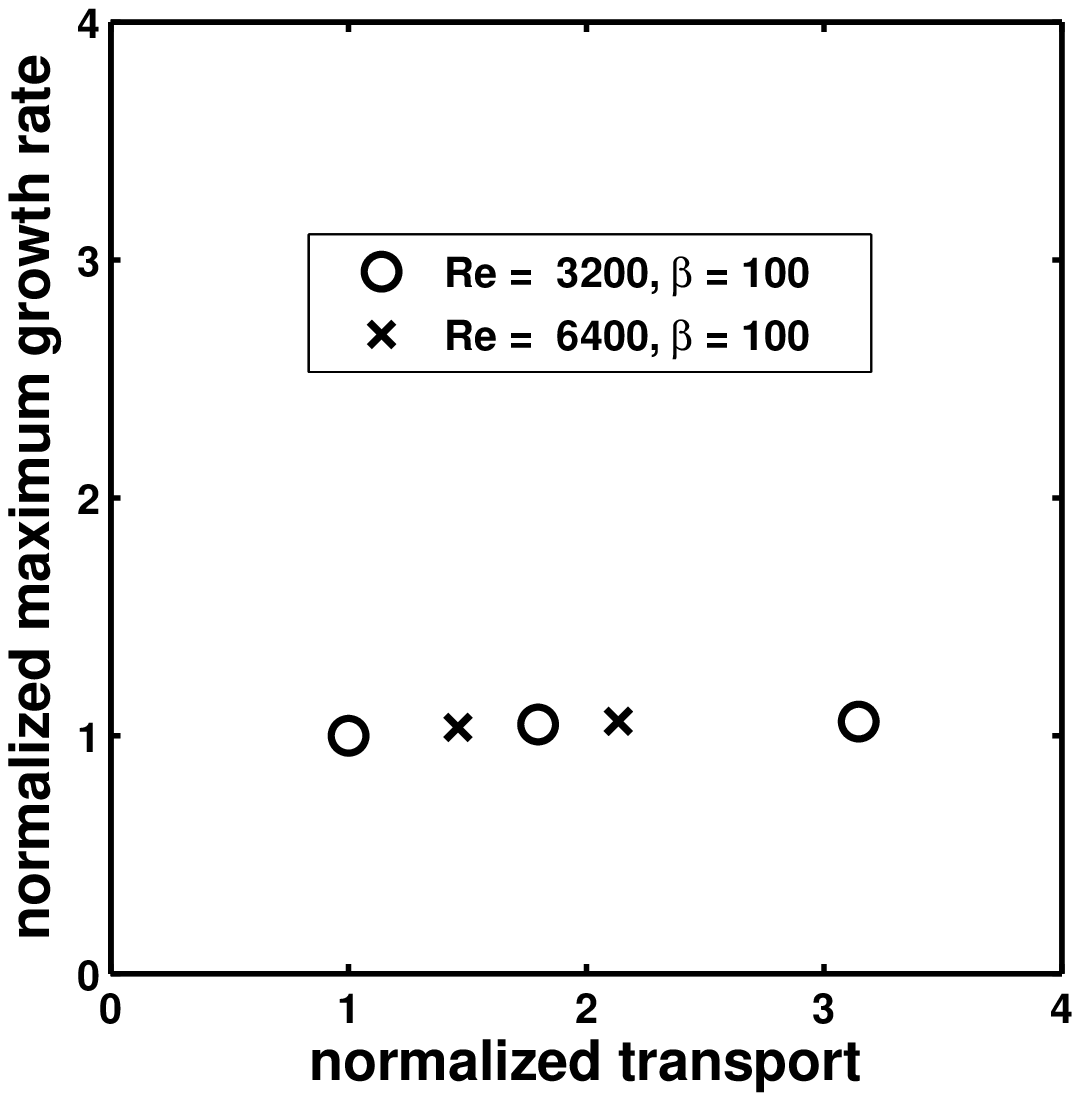}
   \includegraphics[scale=0.5]{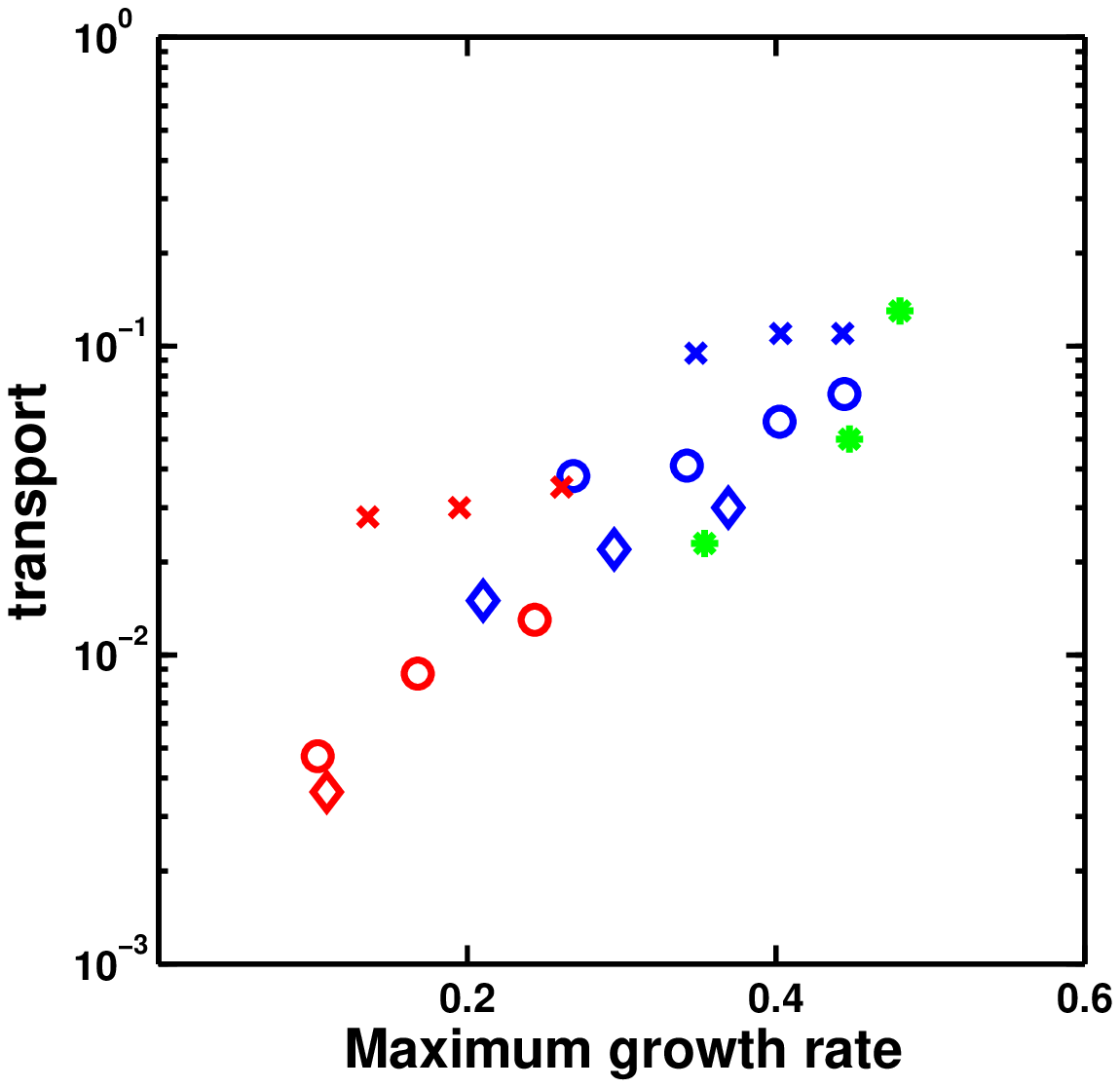}
   \caption{The relation between dimensionless turbulent radial momentum transport
$\alpha$ and the dimensionless growth rate $\sigma/S$. There is no correlation at
the  largest $\beta$ and $Re$ (left) while there is one for weaker field strengths (right). In the left panel, 
the transport and growth rate of the $Re=3200$, $Pm=1/4$ run have been used to normalize the data. Black:
$\beta=10^2$; blue: $\beta=10^3$; red: $\beta=10^4$;
$\diamondsuit$: $Pm=1/4$; $\circ$: $Pm=1$; $\times$: $Pm=4$. The
three resolved runs are also shown (green stars).}\label{tpsig}
\end{figure*}

On the contrary, the spread in Reynolds number for $Pm \le 1$ is
substantial, and systematic. Such a spread was not detected in our
earlier investigation, due to the larger fluctuations in transport
related to the box aspect ratio, as discussed earlier. In fact,
this dispersion seems to be an effect of the magnetic Reynolds
number. To illustrate this point, the transport is represented on
Fig.~\ref{transpReRm} as a function of $Rm$ (left panel) and $Re$
(right panel), for $Pm \le 1$; the colors describe different field
strengths ($\beta=10^2$ to $10^4$ from top to bottom). The
statistics in the number of points at any given $Re$ or $Rm$ is
rather low; however, it appears quite clearly that the dispersion
of the points at any given Reynolds number is substantially larger
in $Re$ (with varying $Rm$) than in $Rm$ (with varying $Re$). 
The largest Reynolds number data strongly support this conclusion.
Furthermore, the fits\footnote{The $Re=20000$ data points have not been included in this fit to make the comparison between the two dependences in the same conditions.} of
the transport as a function of $Rm$ indicate the $Rm$ dependence
of the transport for $Pm \le 1$ is very similar to its $Pm$
dependence as shown on Fig.~\ref{transp-pm}. This strongly
suggests that the $Pm$ dependence observed on this figure is in
fact mostly a $Rm$ dependence for $Pm \le 1$. Including the $Pm=4$
data destroys this correlation, which strengthens the idea that
there are two regimes, depending on the Prandtl number (a feature
that may be related to the existence of a transition around $Pm=2$
in zero net flux shearing box simulations). The relevant results of \cite{SH09}; although less detailed, are consistent with these findings (see their Fig.~7).

\section{Role of channel and parasitic modes}\label{linphys}

\subsection{Linear physics and turbulent transport}

\cite{LL07} concluded that there was no direct connection between
the Prandtl dependence of MRI-driven turbulent transport and the linear growth
rate of channel modes. However,
the transport was substantially less precisely determined than in
the present investigation, and the issue is worth reinvestigating.

To this effect, one needs to compute the linear instability growth
rates for the various runs we have performed. We focus on the
growth rate of the channel mode; for one thing, it is a guideline
for the stability behavior of all unstable modes, and for another,
there is a particular interest in the behavior of this specific
mode. The resulting growth rates are numerically computed from the
dispersion relation recalled in section \ref{rel-disp} and represented
in Fig.~\ref{growthrate-comp}. The important features to note from
this figure are the following:

\begin{itemize}
  \item The largest growth rate in the continuum limit is always
  represented or very nearly represented in the simulation, in spite of
  the discreteness imposed by the vertical boundary condition
  (the corresponding modes are shown by crosses on the figure).
  This ensures that the fastest growing channel mode is always
  present in our simulations.
  \item For the largest field strength ($\beta=100$), the growth
  rates of all our simulations is very close to the ideal MHD
  limit ($\sigma=S/2$). This is especially true
  of the two largest Reynolds numbers ($Re=3200$ and $Re=6400$).
  Note in particular for these runs that the growth rate
  changes by at most $10\%$ while the transport varies by a
  factor of more than $3$ (see left panel of Fig..~\ref{tpsig}).
  As a consequence \textit{the existence of the Prandtl and magnetic Reynolds
  number dependences of the transport 
  is not related to the variation of
  the growth rate with physical dissipation}.
  \item For $\beta=1000$, and for the two largest Prandtl numbers
  ($Pm=1$ and $Pm=4$) and largest Reynolds numbers ($Re=3200$ and
 $Re=6400$), the growth rate is again almost independent of the
 dissipation. As there is only one captured channel mode for the $\beta=100$
  \textit{vs} four or five for $\beta=1000$, \textit{the
  $Pm,Rm$ dependence is not related to the
  number of captured channel modes in the simulation}.
\end{itemize}

There is a sharp behavior difference of the transport with respect to the linear MRI growth rate between the nearly 
ideal simulations performed at $\beta =100$ and the lower $\beta$ values, as is apparent when comparing the left 
and right panels of Fig.~\ref{tpsig}. In fact, the whole spectrum of possible behaviors is covered here, from 
substantial  variation of the transport with very weakly varying growth rates (left panel), or conversely substantial 
variation of the growth rate while the transport is only weakly or mildly changing ($Pm=4$ data on the right 
panel) and intermediate situations where the two are more strongly correlated (the other data points on the 
right panel). Therefore, although this situation is more contrasted than indicated in \cite{LL07}, the conclusion stands: there 
is no clear and direct relation between the effects of dissipation on linear physics and the nonlinear turbulent 
saturation mechanism.

\subsection{Quasi-linear physics and turbulent transport:}

As indicated in the introduction and in section \ref{chan-paras},
it is often assumed that the channel mode is responsible for most
of the MRI-driven turbulent transport in the presence of a net
vertical magnetic field. In particular \cite{PG09} argue that the
saturation of the channel mode by its parasites may help us to
understand the transport. In this section, we make use of our
numerical results to investigate these issues.

\begin{figure*}[!ht]
   \centering
   \includegraphics[scale=0.5]{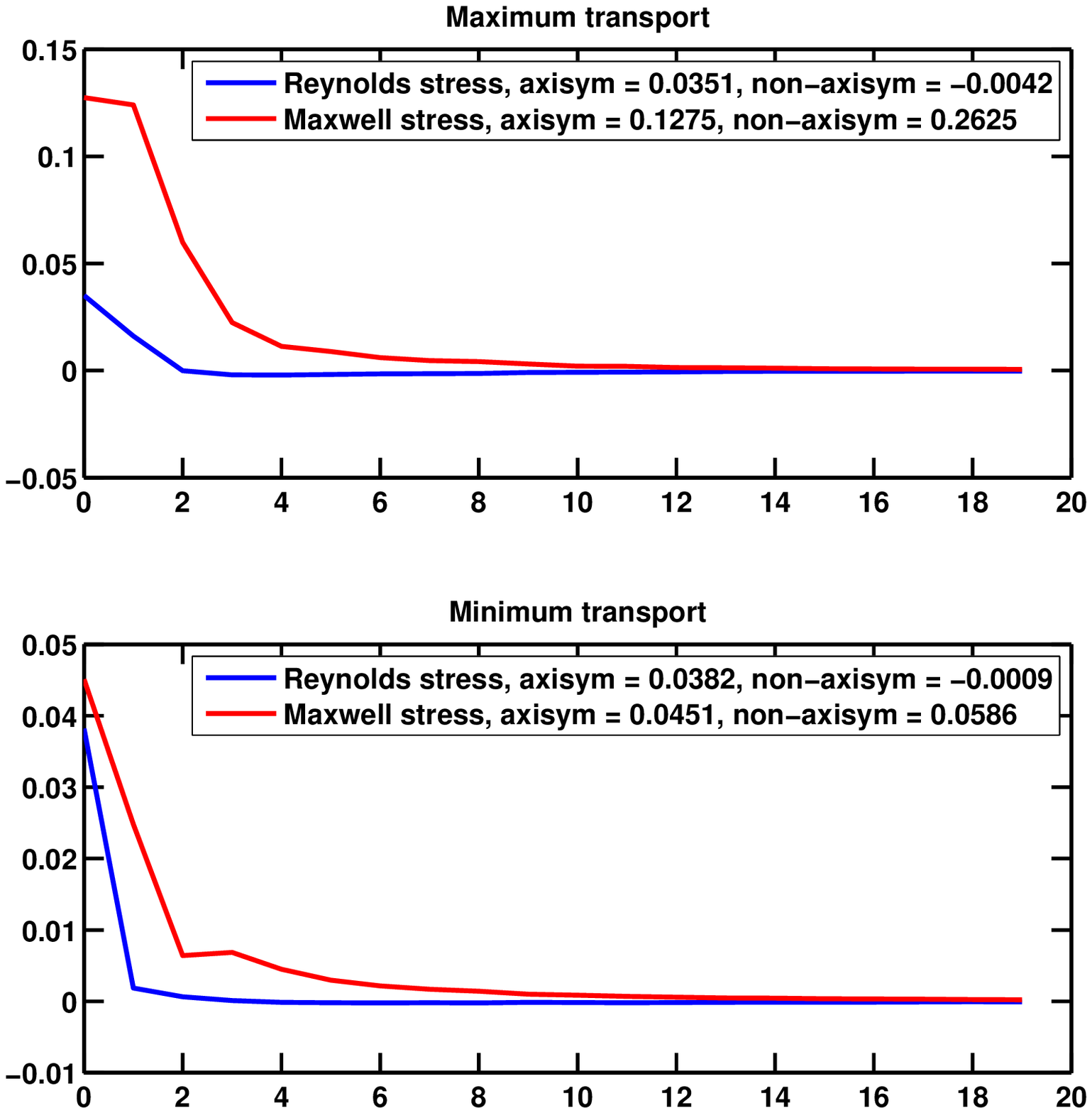}\quad \quad \quad
   \includegraphics[scale=0.5]{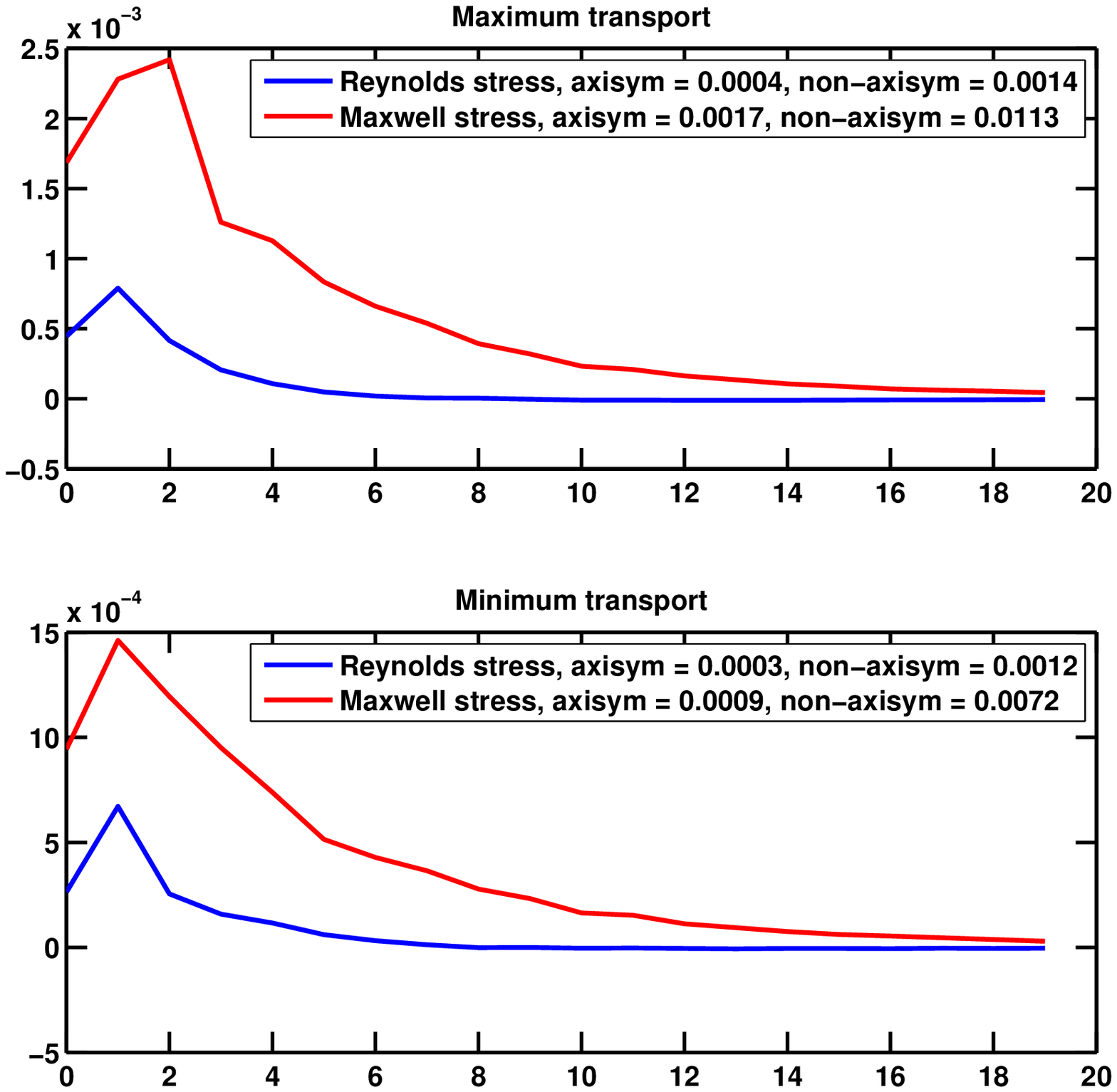}
   \caption{One-dimensional spectrum of the transport, in the azimuthal direction,
   as a function of the wavenumber $n_\phi$ such that $k_\phi=2\pi n_\phi/L_\phi$.
   $\beta=10^2$ for the two left figures, and $\beta=10^4$ for
   the two right ones. The upper plots correspond in each case
   to a typical transport maximum in the time history, while the
   lower ones describe a typical minimum. Both the Reynolds and
   Maxwell stress are shown. The transport due to non-axisymmetric
   modes always dominates the transport due to the axisymmetric
   ones (channel modes). See text.}\label{tp-spec}%
\end{figure*}

We first disprove the first statement: the channel mode does not
clearly dominate transport, especially at the maxima of the
transport, as should in the scheme depicted on Fig.~\ref{satproc}. This is shown
explicitly in Fig.~\ref{tp-spec}. A one-dimensional Fourier
transform of both the Maxwell and Reynolds stress along the
azimuthal direction is shown in this figure. These transforms are
performed at a typical minimum and maximum of transport in the
transport history of two of our runs, namely the runs with
$Re=Rm=6400$ at $\beta=100$ and $\beta=10^4$, respectively. The
transform is shown in terms of the wavenumber $n=k_\phi
L_\phi/2\pi$; $n=0$ corresponds to the transport due to the
axisymmetric modes (channel modes) present in the simulations, for
the various captured $k_z$. The other wavenumbers ($n\ne 0$)
correspond to non-axisymmetric modes. Channel modes dominate over
other individual wavenumbers for $\beta=100$. However, the
cumulated contribution of non-axisymmetric modes is dominant by a
factor $\lesssim 2$ at the maximum of transport for this $\beta$,
and does not exceed the non-axisymmetric contribution by more than
$\sim 30\%$ at minimum transport. For $\beta=10^4$, the cumulated
non-axisymmetric are always dominant by a factor $\sim 6$. For
$\beta=10^3$ (not shown), the situation is somewhat intermediate
between the two extreme values presented in the figure, but the
non-axisymmetric contribution is again always dominant by a factor
several.

In summary, except for minimum transport for the largest field
strength where axisymmetric and non-axisymmetric contributions to
the transport are comparable, the non-axisymmetric contribution is
always larger by a factor $\sim 5$. As a result, it seems
necessary at least to include non-axisymmetric modes in the
process sketched on Fig.~\ref{satproc}.

Furthermore, the analysis of \cite{PG09} is inconsistent with our
results, both from a qualitative and quantitative point of view.
Our simulations always capture the fastest growing channel mode,
or very nearly so (see Fig.~\ref{growthrate-comp}). Due to the
choice of aspect ratio, the fastest growing parasitic modes are
always captured as well. Finally, for $\beta=100$ and $Re=3200$
and $6400$ (even $1600$ for the two larger $Pm$ values), the
variation of the transport with $Pm$ is quite substantial, whereas
\cite{PG09} conclude that in these conditions, the transport is
independent of dissipation (see the left plot in Fig.~2 of their
paper; the three Reynolds numbers values mentioned correspond to
the three largest Elsasser numbers of their graph). 

%Note also that, in the \cite{PG09} picture, the channel mode amplitude 
%at saturation follows from the equality of the parasitic and channel modes 
%growth rates, i.e. when $(B_{ch}/B_0)S \simeq S$ in the dissipationless limit. 
%As the transport $\alpha_{ch}$ due to the most unstable channel mode is related to its amplitude:
%$\alpha_{ch}=(4/5)(B_{ch}/B_0)^2(1/\beta)$ (see \citealt{GX94} or \citealt{PC08}), so the transport 
%should scale with $\beta^{-1}$ and not $\beta^{-1/2}$ quoted in the literature.

\section{Discussion and conclusion}\label{discussion}

Perhaps the most significant new result of this work, disclosed on
Figs.~\ref{transp-pm} and \ref{transpReRm}, is the existence of a
double regime separated by a critical magnetic Prandtl number
$Pm_c\sim 1$. For $Pm < Pm_c$, at a given field strength, the
transport correlates mostly with $Rm$; for $Pm > Pm_c$, the
transport seems to depend mostly on $Pm$ and only weakly on either $Re$ or $Rm$ (once $Re \gtrsim 10^3$), although 
a larger number of $Pm$ values need to be probed on this issue. It is
tempting to assume $Pm_c\simeq 2$, as this is the critical value
for the zero mean field problem, but this identification requires
further work to be substantiated. The identification of this
double dissipation regime was made possible by the increased
accuracy, with respect to our previous work, in the determination
of the transport averages. In the small Prandtl regime, in contrast to 
the large one, our most resolved simulations show no sign of convergence with respect to
dissipation, although values of $Rm$ up to 20000 have been reached.

The role of linear physics and parasitic modes on transport
properties has also been investigated, and the major results on
these questions can be summarized as follows:

\begin{itemize}
  \item The existence of the dependence of turbulent transport on
  dissipation is not related to the role of dissipation on
  linear modes growth rates.
  \item The existence of the correlation of transport with
  dissipation is not related to the number of channel modes
  captured in the simulations.
  \item Saturation of the channel modes growth by parasitic modes
  is not the responsible for correlation of transport with
  dissipation. This is hardly surprising as both the channel and
  parasitic modes are large scale modes, that are expected to be little
  affected by dissipation, whereas the trends of the transport with dissipation are
  substantial.
  \item Furthermore, the transport is usually not dominated by
  axisymmetric modes; these modes contribute at most at the same
  level as non-axisymmetric modes for the strongest fields
  investigated, and have negligible contributions for the weakest
  fields.
\end{itemize}

Note that all these results were obtained while the fastest
growing channel and parasitic modes were always captured, so that
they do not depend on limitations on this front.

It is worth pointing out that our results do not totally
disqualify a saturation of the unstable linear modes by the
parasitic modes; only the relevance of this process to the
relation between transport and physical dissipation has been disproved.
%For instance, the number of unstable modes with growth rates $\sim
%S/2$ scales roughly like $\beta^{1/2}$. As one might expect that a
%fraction of order unity of these modes is at any given time within
%some factor of order unity of saturation, the transport predicted
%by a \cite{PG09} type of argument would then indeed scale like
%$\beta^{-1/2}$. However, such a scheme, if relevant, will require
%substantially more numerical and analytic work to be confirmed, in
%particular on the question of reproducing the correct saturation
%magnitudes (note, as suggested below, that the saturated transport
%may be underestimated in the simulations if the small-scales do
%not react on the large ones in the limit of negligible
%dissipation).

The role of the Prandtl number disclosed in this investigation
makes more physical sense than a direct dependence of the
transport on $Pm$. Indeed, it is very likely that the critical
value $Pm_c$ relates to the switching of the magnetic and kinetic
dissipations scales $k_\eta$ and $k_\nu$ in Fourier space: for $Pm
< Pm_c$ (resp.\ $Pm > Pm_c$), $k_\eta < k_\nu$ (resp.\ $k_\eta >
k_\nu$). For $Pm \ll 1$, $k_\eta \ll k_\nu$, the flow at scales $k
>$ or $\gg k_\eta$ in Fourier space is therefore purely
hydrodynamic. As information flows from small $k$ to large $k$ in
this scale range, the flow should be independent of $Re$, so that in
this regime $\alpha=\alpha(\beta,Rm)$ is expected. The transport
may even become independent of $Rm$ at large enough $Rm$ in the
absence of backreaction of the small scales on the large ones, and
if energy exchanges are not strongly nonlocal in Fourier space. We merely note
here that in the context of homogeneous, isotropic, incompressible
MHD turbulence, all energy transfers are direct (from small $k$ to
large $k$); exchanges between magnetic and kinetic energy are nonlocal 
in Fourier space  \citep{AMP07}, although this nonlocality seems to be of finite (albeit large) extent \citep{AE09}.
The  situation is more complex in the large Prandtl regime, as non-local transfers and small scale dynamo action might take place at scales smaller than the viscous dissipation scale, and further investigations are required to characterize the properties of the small dissipation limit, in particular concerning the locality of transfers.

The behavior of MRI-driven turbulence with respect to these
various issues will be reported elsewhere, through the analysis of
energy transfers in Fourier space. More generally, investigating
the possible independence of transport with respect to dissipation
in the $Pm \ll Pm_c$ or $Pm \gg Pm_c$ limits requires to achieve
a double scale separation: one must first separate in Fourier
space the smallest injection scales (the scales which contribute to the
Reynolds and Maxwell stresses) from the largest dissipation
scales, and then the two dissipation scales themselves. This is
extremely demanding in terms of resolution. Our simulations in the
small Prandtl regime are consistent with a separation of
dissipation scales nearly achieved. As a first step at small Prandtl
(the most critical regime for YSO disks), one can achieve one or
the other of the two scales separation. Such investigations are
underway and will be reported elsewhere.

These questions are critical for
astrophysical disk dynamics, where the Reynolds numbers are always very large ($Re \gtrsim 10^{10}$, while the Prandtl number spans very small ($\lesssim 10^{-5}$, YSO disks) to very large values ($\gtrsim 10^{5}$, AGN disks). Addressing these issues requires to overcome a number of
limitations in the simulations. Besides the questions of scale
separation and nonlocality of transfers in Fourier space
just mentioned, the role of compressibility and vertical
stratification on these results must be quantified. These
questions can certainly be investigated in shearing boxes, but
global simulations are probably still largely out of reach, due to
the resolution demand. Also (and probably as much importantly),
different dissipation regimes must be analyzed in the context of
YSOs, namely the ambipolar and Hall regime, which are known to
have substantial effect on the linear development of the MRI,
while being relevant for large fractions of YSO disks.

\begin{acknowledgements}

The simulations presented in this work have been performed on the
French national supercomputing center at IDRIS. PYL acknowledges
the hospitality of the \textit{Isaac Newton Institute} in
Cambridge where most of the data analysis has been done. GL acknowledges support
from STFC. The authors thank S\'ebastien Fromang and an anonymous referee for their comments on a first version of this work.

\end{acknowledgements}

\bibliographystyle{aa}
\bibliography{14093}

\end{document}